\documentclass{article}%
\usepackage{amsmath}
\usepackage{amsfonts}
\usepackage{amssymb}
\usepackage{graphicx}%
\providecommand{\U}[1]{\protect\rule{.1in}{.1in}}

\begin{document}

\title{Entropic Dynamics: an Inference Approach to Quantum Theory, Time and
Measurement\thanks{Invited paper presented at the EmQM13 Workshop on Emergent
Quantum Mechanics, Austrian Academy of Sciences (October 3--6, 2013, Vienna).
}}
\author{Ariel Caticha\\{\small Department of Physics, University at Albany-SUNY, }\\{\small Albany, NY 12222, USA.}}
\date{}
\maketitle

\begin{abstract}
We review the derivation of quantum theory as an application of entropic
methods of inference. The new contribution in this paper is a streamlined
derivation of the Schr\"{o}dinger equation based on a different choice of
microstates and constraints.

\end{abstract}

\section{An overview}

Quantum mechanics involves probabilities in a fundamental way and, therefore,
it is a theory of inference. But this has not always been clear. The
controversy revolves around the interpretation of the quantum state --- the
wave function. Does it represent the actual real state of the system --- its
\emph{ontic} state --- or does it represent a state of knowledge about the
system --- an \emph{epistemic} state?

Examples of ontic interpretations include, to name a few, Bohm's causal
interpretation, Everett's many-worlds interpretation, and Nelson's stochastic
mechanics \cite{Nelson 1985}. The epistemic interpretations have also had a
number of advocates (for example, \cite{Ballentine 1970}-\cite{Harrigan
Spekkens 2010}) starting, most prominently, with Einstein. The
\textquotedblleft orthodox\textquotedblright\ or Copenhagen interpretation
lies somewhere in between. On one hand, as described in standard textbooks
such as the early classics by Dirac and von Neumann, it regards the quantum
state as a complete and objective specification of the properties of the
system --- an ontic concept that is divorced from the state of knowledge of
any rational agent. On the other hand, the founders of quantum theory ---
Bohr, Heisenberg, Born, etc. --- were keenly aware of the epistemological and
pragmatic elements in quantum mechanics (see e.g., \cite{Stapp 1972}) but,
unfortunately, they wrote at a time when the language, the tools and the rules
of quantitative epistemology --- the Bayesian and entropic methods of
inference --- had not yet been sufficiently developed. As a result they did
not succeed in drawing a sharp line between the ontic and the epistemic and
thereby started a controversy that lingers to this day.

But interpreting quantum theory is not merely a matter of postulating the
mathematical formalism and then appending an interpretation to it. For the
epistemic view of quantum states to be satisfactory it is not sufficient to
state that the probability $\left\vert \psi\right\vert ^{2}$ represents a
state of knowledge; we must also show that changes or updates of the epistemic
$\psi$ --- which include both the unitary time evolution according to the
Schr\"{o}dinger equation and the projection postulate during measurement ---
obey the rules of inference. In a truly epistemic interpretation there is no
logical room for \textquotedblleft quantum\textquotedblright\ probabilities
obeying alternative rules of inference.

Our subject is Entropic Dynamics (ED), a framework in which quantum theory is
formulated as an example of entropic inference \cite{Caticha 2010}. ED differs
from other approaches in several important respects. For example, in the
standard view quantum theory is considered as an extension of classical
mechanics and therefore deviations from causality demand an explanation. In
the entropic view, on the other hand, quantum mechanics is an example of
entropic inference, a scheme designed to handle insufficient information
\cite{Caticha 2012}. From the entropic perspective indeterminism requires no
explanation: uncertainty and probabilities are the norm. It is certainty and
determinism that demand explanations.

ED also differs from other approaches based on information theory. (See
\emph{e.g.}, \cite{Wootters 1981}-\cite{Reginatto 2013}.) In ED\ the laws of
physics are rules for processing information. The information in question
possibly originates and might even find its ultimate justification in some
sub-quantum dynamics that remains to be discovered. However, as we shall see,
once the relevant information has been identified, the remaining details of
any such underlying dynamics turn out to be irrelevant for behavior at the
quantum level. In ED those irrelevant details are ignored from the start ---
which is a significant simplification. The situation is somewhat analogous to
the laws of thermodynamics which also turns out to be largely independent of
microscopic details at the atomic level.

The analogy with thermodynamics has inspired several attempts to explain the
emergence of quantum behavior from specific proposals of a sub-quantum
dynamics with some additional stochastic element. (See \emph{e.g.},
\cite{Nelson 1985}\cite{de la Pena Cetto 1996}-\cite{Grossing 2008}.) In
contrast, ED does not assume any underlying mechanics whether classical,
deterministic, or stochastic. Both quantum dynamics and its classical limit
are derived as examples of entropic inference.

Another difference is that ED naturally leads to an \textquotedblleft
entropic\textquotedblright\ notion of time. Time is introduced as a convenient
book-keeping device to keep track of the accumulation of change. The task is
to develop a model that includes (a) something one might identify as an
\textquotedblleft instant\textquotedblright, (b) a sense in which these
instants can be \textquotedblleft ordered\textquotedblright, (c) a convenient
concept of \textquotedblleft duration\textquotedblright\ measuring the
separation between instants. The welcome new feature is that entropic time is
intrinsically directional. Thus, an arrow of time is generated automatically.

ED offers a new perspective on the notorious problem of measurement (see
\cite{Ballentine 1998}\cite{Schlosshauer 2004}\cite{Jaeger 2009}). Questions
such as \textquotedblleft How can a measurement ever yield a definite
outcome?\textquotedblright\ or \textquotedblleft Are the values of observables
created during the act of measurement?\textquotedblright\ led von Neumann to
postulate a dual mode of wave function evolution, either continuous and
deterministic according to the Schr\"{o}dinger equation, or discontinuous and
stochastic during the measurement process. Once one accepts quantum theory as
a theory of inference the dichotomy between the two modes disappears. Unitary
evolution and discontinuous collapse correspond to two modes of updating
probabilities which, as shown in \cite{Caticha 2006}, are not intrinsically
different; they are special cases within a broader scheme of entropic
inference \cite{Caticha 2012}.

Yet another distinguishing feature is that in ED the positions of particles
have definite values just as they would in classical physics. This implies
that the process of observation is essentially classical and measurements of
position automatically yield definite outcomes. This solves the problem of
measurement because position is the only observable. Indeed, in ED all other
observables such as momentum, energy, and so on, are statistical concepts ---
just like temperature in statistical mechanics. They are not properties of the
particles but of their probability distributions. As shown in \cite{Caticha
2000}\cite{Johnson Caticha 2011} their values are indeed created by the act of
measurement \cite{Nawaz Caticha 2011}. A more detailed treatment is given in
\cite{Johnson 2011} and \cite{Caticha 2012}.

In order to formulate quantum theory as an entropic dynamics --- just as with
any other inference problem --- we must decide which microstates are the
subject of our inference, we must identify the prior probabilities, and we
must identify those constraints that represent the information that is
relevant to our problem. The new contribution in this paper is an entropic
derivation of the Schr\"{o}dinger equation based on a choice of microstates
and constraints that differs and is in some respects more advantageous than
the choice adopted in \cite{Caticha 2010}.

\section{Entropic Dynamics}

In this model we consider particles living in flat three-dimensional space.
The particles have \emph{definite} positions and it is their unknown values
that we wish to infer.\footnote{In this work ED is developed as a model for
the quantum mechanics of particles. The same framework can be deployed to
construct models for the quantum mechanics of fields, in which case it is the
fields that are objectively \textquotedblleft real\textquotedblright\ and have
well-defined albeit unknown values.}

The basic dynamical assumptions are that motion happens and that it is
continuous. Thus, short displacements happen and it is their accumulation that
leads to motion. We do not explain why motion happens but, given the
information that it does, our task is to venture a guess about what to expect.

For simplicity here we will focus on a single particle; the generalization to
$N$ particles is straightforward. For a single particle the configuration
space is $\mathbb{R}^{3}$ with metric $\delta_{ab}$. The particle moves from
an initial $x$ to an unknown $x^{\prime}$. The goal is to to find the
probability distribution $P(x^{\prime}|x)$. To find it maximize the
appropriate (relative) entropy,
\begin{equation}
\mathcal{S}[P,Q]=-\int d^{3}x^{\prime}\,P(x^{\prime}|x)\log\frac{P(x^{\prime
}|x)}{Q(x^{\prime}|x)}~. \label{Sppi}%
\end{equation}
The relevant information is introduced through the prior probability
$Q(x^{\prime}|x)$, which reflects our knowledge about which $x^{\prime}$ to
expect \emph{before} we have any information about the motion, and the
constraints that specify the family of acceptable posteriors $P(x^{\prime}|x)$.

\paragraph*{The prior}

We adopt a prior that represents a state of extreme ignorance. Knowledge of
$x$ tells us nothing about $x^{\prime}$. Such ignorance is represented by a
uniform distribution: $Q(x^{\prime}|x)d^{3}x^{\prime}$ is proportional to the
volume element $d^{3}x^{\prime}$. (The proportionality constant has no effect
on the entropy maximization and can be safely ignored.)

\paragraph*{The constraints on the motion}

The information that motion is continuous is imposed through a constraint. For
a short step let $\vec{x}^{\prime}=\vec{x}+\Delta\vec{x}$. We require that the
expected squared displacement,
\begin{equation}
\left\langle \Delta\vec{x}\cdot\Delta\vec{x}\right\rangle =\kappa~,
\label{kappa}%
\end{equation}
be some small but for now unspecified value $\kappa$, which we take to be
independent of $x$ in order to ref{}lect the translational symmetry of the
configuration space is $\mathbb{R}^{3}$.

If this were the only constraint the resulting motion would be a completely
isotropic diffusion. Clearly some information is still missing. The additional
piece of relevant information, that once particles are set in motion they tend
to persist in it, is expressed by assuming the existence of a
\textquotedblleft potential\textquotedblright\ $\phi(x)$ and imposing that the
expected displacement $\left\langle \Delta x^{a}\right\rangle $ in the
direction of the gradient of $\phi$ is constrained to be%
\begin{equation}
\left\langle \Delta\vec{x}\right\rangle \cdot\vec{\nabla}\phi=\kappa^{\prime}%
\end{equation}
where $\kappa^{\prime}$ is another small but for now unspecified
position-independent constant.

The seemingly ad hoc introduction of a potential $\phi$ will not be justified
--- at least not here. The important point is that we have identified the
information needed for inference. Where this information originates and why it
turns out to be relevant are, of course, interesting questions but their
answers lie elsewhere --- at some deeper level of physics. For the purpose of
inference no further hypotheses need be made.\footnote{In our earlier
development of ED \cite{Caticha 2010} the set of microstates involved the
positions $x$ and some additional mysterious variables that we called $y$. The
present treatment is simpler in that no $y$ variables need be postulated. This
simplification comes at a price. In the $y$-variable model both the potential
$\phi$ and the appearance of its gradient arise naturally without further
assumptions. Thus we have a trade-off. We can simplify the microstates at the
expense of the constraints. A deeper justification for $\phi$, its geometric
significance, and its natural relation to gauge symmetries can be given and
will be discussed elsewhere in the context of particles with spin.}

Having specified the prior and the constraints the ME method takes over.
Varying $P(x^{\prime}|x)$ to maximize $\mathcal{S}[P,Q]$ in (\ref{Sppi})
subject to the two constraints plus normalization gives
\begin{equation}
P(x^{\prime}|x)=\frac{1}{\zeta}\exp[-\frac{1}{2}\alpha\,\Delta\vec{x}%
\cdot\Delta\vec{x}+\alpha^{\prime}\Delta\vec{x}\cdot\vec{\nabla}\phi]~,
\label{Prob xp/x}%
\end{equation}
where $\zeta$ is a normalization constant,
\begin{equation}
\zeta(x,\alpha,\alpha^{\prime})=\int d^{3}x^{\prime}\,e^{-\frac{1}{2}%
\alpha\,\Delta\vec{x}\cdot\Delta\vec{x}+\alpha^{\prime}\Delta\vec{x}\cdot
\vec{\nabla}\phi}~.
\end{equation}
The Lagrange multipliers $\alpha$ and $\alpha^{\prime}$ are determined in the
standard way,
\begin{equation}
\partial\log\zeta/\partial\alpha=-\kappa/2\quad\text{and\quad}\partial
\log\zeta/\partial\alpha^{\prime}=-\kappa^{\prime}~.
\end{equation}
Since both the function $\phi$ and the constant $\kappa^{\prime}$ are so far
unspecified, so is the multiplier $\alpha^{\prime}$. Without loss of
generality, we can absorb $\alpha^{\prime}$ into $\phi$, $\alpha^{\prime}%
\phi\rightarrow\phi$, which amounts to setting $\alpha^{\prime}=1$.

Eq.(\ref{Prob xp/x}) for $P(x^{\prime}|x)$ shows that short steps are obtained
for large $\alpha$ and that they happen in essentially random directions with
a small anisotropic bias along the gradient of $\phi$. The distribution
$P(x^{\prime}|x)$ is Gaussian and is conveniently written as
\begin{equation}
P(x^{\prime}|x)\propto\exp\left[  -\frac{\alpha}{2}\left(  \Delta\vec
{x}-\left\langle \Delta\vec{x}\right\rangle \right)  ^{2}\right]  ~.
\label{Prob xp/x b}%
\end{equation}
The displacement $\Delta\vec{x}=\Delta\bar{x}+\Delta\vec{w}$ can be expressed
as the expected drift plus a fluctuation%
\begin{equation}
\left\langle \Delta\vec{x}\right\rangle =\Delta\bar{x}=\frac{1}{\alpha}%
\vec{\nabla}\phi~, \label{ED drift}%
\end{equation}%
\begin{equation}
\left\langle \Delta w^{a}\right\rangle =0\quad\text{and}\quad\left\langle
\Delta w^{a}\Delta w^{b}\right\rangle =\frac{1}{\alpha}\delta^{ab}~.
\label{ED fluctuations}%
\end{equation}
As $\alpha\rightarrow\infty$ the fluctuations become dominant: the drift
$\Delta\bar{x}\sim\alpha^{-1}$ while $\Delta\vec{w}\sim\alpha^{-1/2}$. This
implies that, as in Brownian motion, the trajectory is continuous but not
differentiable. Here we see the roots of the uncertainty principle: a particle
has a definite position but its velocity, the tangent to the trajectory, is
completely undefined.

\section{Entropic time}

The foundation of all notions of time is dynamics. In ED time is introduced as
a book-keeping device to keep track to the accumulation of small changes.

\subsection{An ordered sequence of instants}

In ED, at least for infinitesimally short steps, change is given by the
transition probability $P(x^{\prime}|x)$ in eq.(\ref{Prob xp/x b}). The $n$th
step takes us from $x=x_{n-1}$ to $x^{\prime}=x_{n}$. Using the product rule
for the joint probability, $P(x_{n},x_{n-1})=P(x_{n}|x_{n-1})P(x_{n-1})$, and
integrating over $x_{n-1}$, we get
\begin{equation}
P(x_{n})=\int d^{3}x_{n-1}\,P(x_{n}|x_{n-1})P(x_{n-1})~. \label{CK a}%
\end{equation}
This equation is a direct consequence of the laws of probability. However, if
$P(x_{n-1})$ happens to be the probability of different values of $x_{n-1} $
\emph{at a given instant labelled }$t$, then we will interpret $P(x_{n})$ as
the probability of values of $x_{n}$ \emph{at the \textquotedblleft
later\textquotedblright\ instant }$t^{\prime}=t+\Delta t$. Accordingly, we
write $P(x_{n-1})=\rho(x,t)$ and $P(x_{n})=\rho(x^{\prime},t^{\prime})$ so
that
\begin{equation}
\rho(x^{\prime},t^{\prime})=\int d^{3}x\,P(x^{\prime}|x)\rho(x,t) \label{CK b}%
\end{equation}
Nothing in the laws of probability that led to eq.(\ref{CK a}) forces this
interpretation on us --- this is an independent assumption about what
constitutes time in our model. We use eq.(\ref{CK b}) to define what we mean
by an instant:\emph{\ if the distribution }$\rho(x,t)$\emph{\ refers to one
instant }$t$\emph{, then the distribution }$\rho(x^{\prime},t^{\prime}%
)$\emph{\ defines what we mean by the \textquotedblleft next\textquotedblright%
\ instant }$t^{\prime}=t+\Delta t$. Thus, eq.(\ref{CK b}) allows
\emph{entropic time} to be constructed one instant after another.

We can phrase this idea somewhat differently. Once we have decided on the
relevant information necessary for predicting future behavior we can imagine
all that information codified into an \textquotedblleft
instant\textquotedblright. Thus, we define instants so that\emph{\ given the
present the future is independent of the past.}\footnote{An equation such as
(\ref{CK b}) is commonly employed to define Markovian behavior in which case
it is known as the Chapman-Kolmogorov equation. Markovian processes are such
that specifying the state of the system at time $t$ is sufficient to fully
determine its state after time $t$ --- no additional information about the
past is needed. We make no Markovian assumptions. We are concerned with a
different problem. We do not use (\ref{CK b}) to define Markovian processes;
we use it to define time.}\emph{\ }

\subsection{The arrow of entropic time}

The notion of time as constructed according to eq.(\ref{CK b}) is remarkable
in that it incorporates an intrinsic directionality: there is an absolute
sense in which $\rho(x,t)$\ is prior and $\rho(x^{\prime},t^{\prime})$\ is posterior.

Suppose we wanted to find a time-reversed evolution. We would write
\begin{equation}
\rho(x,t)=%
{\textstyle\int}
d^{3}x^{\prime}\,P(x|x^{\prime})\rho(x^{\prime},t^{\prime})\,.
\end{equation}
This is perfectly legitimate but in order to be correct $P(x|x^{\prime})$
cannot be obtained from eq.(\ref{Prob xp/x b}) by merely exchanging $x$ and
$x^{\prime}$. According to the rules of probability theory $P(x|x^{\prime})$
is related to eq.(\ref{Prob xp/x b}) by Bayes' theorem,
\begin{equation}
P(x|x^{\prime})=\frac{P(x)}{P(x^{\prime})}P(x^{\prime}|x)~.
\end{equation}
In other words, one of the two transition probabilities, either $P(x^{\prime
}|x)$ or $P(x|x^{\prime})$, \emph{but not both}, can be given by the maximum
entropy distribution eq.(\ref{Prob xp/x b}). The other is related to it by
Bayes' theorem. There is no symmetry between the inferential past and the
inferential future because there is no symmetry between priors and posteriors.

The puzzle of the arrow of time has a long history (see \emph{e.g.}
\cite{Price 1996}\cite{Zeh 2002}). The standard question has been how can an
arrow of time be derived from underlying laws of nature that are symmetric? ED
offers a new perspective. The asymmetry is the inevitable consequence of
entropic inference. From the point of view of ED the challenge does not
consist in explaining the arrow of time, but rather in explaining how it comes
about that despite the arrow of time some laws of physics turn out to be
reversible. Indeed, even when the derived laws of physics -- in our case, the
Schr\"{o}dinger equation -- turns out to be fully time-reversible,
\emph{entropic time itself only f{}lows forward}.

\subsection{Duration: a convenient time scale}

Having introduced the notion of successive instants we now have to specify the
interval $\Delta t$ between them. This amounts to specifying the multiplier
$\alpha(x,t)$ in terms of $\Delta t$.

\emph{Time is defined so that motion looks simple.} For large $\alpha$ the
dynamics is dominated by the f{}luctuations $\Delta w$. In order that the
f{}luctuations $\left\langle \Delta w^{a}\Delta w^{b}\right\rangle $ ref{}lect
the symmetry of translations in space and time --- a Newtonian time that flows
\textquotedblleft equably\textquotedblright\ everywhere and everywhen --- we
choose $\alpha$ to be independent of $x$ and $t$, $\alpha(x,t)=C/\Delta t$,
where $C$ is some constant.

The extension of ED to several non-identical particles is not our subject here
but a quick remark is useful. The extension is achieved by introducing
separate constraints, eq.(\ref{kappa}), for each particle, each with its own
$\kappa_{i}$, and each with its own multiplier $\alpha_{i}=C_{i}/\Delta t$. It
is convenient to write each of these multipliers $\alpha_{i}$ as $\alpha
_{i}=m_{i}/\hbar\Delta t$ in terms of a particle-specific constant $m_{i}$ and
an overall constant $\hbar$ which fixes the units of the $m_{i}$s relative to
the units of time. Thus
\begin{equation}
\alpha=\frac{m}{\hbar\Delta t}~.
\end{equation}
With this choice of the multiplier $\alpha$ the dynamics is indeed simple:
$P(x^{\prime}|x)$ in\ (\ref{Prob xp/x b}) becomes a standard Wiener process.
The displacement is
\begin{equation}
\Delta\vec{x}=\vec{b}\Delta t+\Delta\vec{w}~, \label{Delta x}%
\end{equation}
where $b^{a}(x)$ is the drift velocity,
\begin{equation}
\langle\Delta\vec{x}\rangle=\vec{b}\Delta t\quad\text{with}\quad\vec{b}%
=\frac{\hbar}{m}\vec{\nabla}\phi~, \label{drift velocity}%
\end{equation}
and $\Delta w^{a}$ is a f{}luctuation,
\begin{equation}
\left\langle \Delta w^{a}\right\rangle =0\quad\text{and}\quad\langle\Delta
w^{a}\Delta w^{b}\rangle=\frac{\hbar}{m}\Delta t\,\delta^{ab}~. \label{fluc}%
\end{equation}
The formal similarity to Nelson's stochastic mechanics \cite{Nelson 1985} is
evident but the interpretations are completely different.

Two remarks are in order: one on the nature of clocks and another on the
nature of mass.

\paragraph*{On clocks:}

Time is defined so that motion looks simple. In Newtonian mechanics the
prototype of a clock is the free particle. Time is defined so that the free
particle moves equal distances in equal times. In ED the prototype of a clock
is a free particle too. (For sufficiently short times all particles are free.)
And time is defined so that the particle undergoes equal fluctuations in equal times.

\paragraph*{On mass:}

The particle-specific constant $m$ will, of course, be called `mass' and
eq.(\ref{fluc}) provides its interpretation: mass is an inverse measure of fluctuations.

\section{Accumulating changes: the Fokker-Planck equation}

Equation is an integral equation for the evolution of the distribution
$\rho(x,t)$. As is well-known from diffusion theory \cite{Caticha 2012} it can
be written in differential form as a Fokker-Planck equation (FP),
\begin{equation}
\partial_{t}\rho=-\vec{\nabla}\cdot(\rho\vec{b})+\frac{\hbar}{2m}\nabla
^{2}\rho~, \label{FP a}%
\end{equation}
which can itself be rewritten as a continuity equation,
\begin{equation}
\partial_{t}\rho=-\vec{\nabla}\cdot\left(  \rho\vec{v}\right)  ~. \label{FP b}%
\end{equation}
The velocity $\vec{v}$ of the probability flow or \emph{current velocity} is%
\begin{equation}
\vec{v}=\vec{b}+\vec{u}\quad\text{where}\quad\vec{u}=-\frac{\hbar}{m}%
\vec{\nabla}\log\rho^{1/2}~, \label{osmo}%
\end{equation}
the \emph{osmotic velocity}, represents the tendency for probability to flow
down the density gradient.

Since both $\vec{b}$ and $\vec{u}$ are gradients, it follows that the current
velocity is a gradient too,
\begin{equation}
\vec{v}=\frac{\hbar}{m}\vec{\nabla}\Phi\quad\text{where}\quad\Phi=\phi
-\log\rho^{1/2}~. \label{curr}%
\end{equation}

With these results ED reaches a certain level of completion: We figured out
what small changes to expect --- they are given by $P(x^{\prime}|x)$ --- and
time was introduced to keep track of how these small changes accumulate; the
net result is diffusion according to the FP equation.

But quantum mechanics is not a \emph{standard} diffusion. The discussion so
far has led us to the density $\rho(x,t)$ as the important dynamical object
but to construct a wave function, $\Psi=\rho^{1/2}e^{i\Phi}$, we need a second
degree of freedom, the phase $\Phi$. The problem is that as long as the
potential $\phi$ is externally prescribed the function $\Phi$ in
eq.(\ref{curr}) does not represent an independent degree of freedom. The
natural solution is to relax this constraint and allow $\phi$ (or equivalently
$\Phi$) to participate in the dynamics. Thus the dynamics will consist of the
coupled evolution of $\rho(x,t)$ and $\Phi(x,t)$.

\section{Non-dissipative diffusion}

To specify the dynamics we follow \cite{Nelson 1979} and impose that the
dynamics be non-dissipative, that is, we require the conservation of a certain
functional $E[\rho,\phi]$ which will be called \textquotedblleft
energy\textquotedblright.

At first sight it might appear that imposing that some energy\ $E[\rho,\phi
]$\ be conserved is natural because it agrees with our classical
preconceptions of what physics ought to be like. But classical intuitions are
not a good guide here. In the more sophisticated approaches to physics energy
is taken to be whatever happens to be conserved as a result of invariance
under translations in time. But our dynamics has hardly been defined yet;
what, then, is $E$\ and why should it be conserved in the first place?
Furthermore, if we go back to eq.(\ref{Delta x}) we see that it is the kind of
equation (a Langevin equation) that characterizes a Brownian motion in the
limit of \emph{infinite} friction. Thus, the explanation of quantum theory in
terms of a sub-quantum classical mechanics would require that particles be
subjected to infinite friction while suffering zero dissipation at the same
time. Such a strange sub-quantum mechanics could hardly be called `classical'.

The energy functional $E[\rho,\phi]$ is chosen to be the expectation of a
\emph{local} \textquotedblleft energy\textquotedblright\ function
$\varepsilon(x,t)$, that is,
\begin{equation}
E[\rho,\phi]=\int d^{3}x\,\rho(x,t)\,\varepsilon(x,t)~, \label{energy a}%
\end{equation}
where $\varepsilon(x,t)$ depends on $\rho(x,t)$ and $\phi(x,t)$ and their
derivatives.\footnote{In an energy eigenstate the local energy $\varepsilon
(x,t)$ is uniform in space and constant in time.} The local energy appropriate
to the non-relativistic regime is
\begin{equation}
\varepsilon(x,t)=\frac{1}{2}mv^{2}+\frac{1}{2}mu^{2}+V(x)~, \label{energy b}%
\end{equation}
where the scalar function $V(x)$ represents an additional \textquotedblleft
potential\textquotedblright\ energy. The justification of $\varepsilon$ is to
be found in deeper-level physics but we can note that $\varepsilon$ is tightly
constrained by requiring invariance under time reversal ($\vec{v}%
\rightarrow-\vec{v}$ and $\vec{u}\rightarrow\vec{u}\,$) and the low velocity
regime \cite{Smolin 2006}\cite{Caticha 2010}.

Using eqs.(\ref{osmo}) and (\ref{curr}) the energy $E$ is
\begin{equation}
E=\int d^{3}x\,\rho\left(  \frac{\hbar^{2}}{2m}(\vec{\nabla}\Phi)^{2}%
+\frac{\hbar^{2}}{2m}(\vec{\nabla}\log\rho^{1/2})^{2}+V\right)  ~
\end{equation}
so that, after some algebra \cite{Caticha 2010},
\begin{equation}
\frac{dE}{dt}=\int d^{3}x\,\dot{\rho}\left(  \hbar\dot{\Phi}+\frac{\hbar^{2}%
}{2m}(\vec{\nabla}\Phi)^{2}+V-\frac{\hbar^{2}}{2m}\frac{\nabla^{2}\rho^{1/2}%
}{\rho^{1/2}}\right)
\end{equation}
We impose that $\dot{E}=0$ for spatially arbitrary choices of the initial
conditions $\rho$ and $\Phi$, that is, at the initial $t_{0}$ we ought to be
able to change $\rho$ and $\Phi$ independently at different locations and
still get $\dot{E}=0$. This implies the integrand should vanish at the initial
$t_{0}$. But any arbitrary time $t$ can be taken as the initial time for
evolution into the future. Therefore for all $t$ we require that
\begin{equation}
\hbar\dot{\Phi}+\frac{\hbar^{2}}{2m}(\vec{\nabla}\Phi)^{2}+V-\frac{\hbar^{2}%
}{2m}\frac{\nabla^{2}\rho^{1/2}}{\rho^{1/2}}=0~, \label{SEb}%
\end{equation}
which is the quantum version of the Hamilton-Jacobi equation. Equations
(\ref{SEb}) and the FP equation,
\begin{equation}
\dot{\rho}=-\vec{\nabla}\cdot\left(  \rho\vec{v}\right)  =-\frac{\hbar}{m}%
\vec{\nabla}\cdot\left(  \rho\vec{\nabla}\Phi\right)  ~ \label{SEa}%
\end{equation}
are the coupled dynamical equations we seek.

These two real equations can be written as a single complex equation by
combining $\rho$ and $\Phi$ into a complex function $\Psi=\rho^{1/2}\exp
(i\Phi)$. Computing the time derivative $\dot{\Psi}$ and using eqs.(\ref{SEb})
and (\ref{SEa}) leads to the Schr\"{o}dinger equation,
\begin{equation}
i\hbar\frac{\partial\Psi}{\partial t}=-\frac{\hbar^{2}}{2m}\nabla^{2}%
\Psi+V\Psi~. \label{SE}%
\end{equation}
Earlier we had introduced $m$ as a particle-specific constant that measures
fluctuations, $\hbar$ as a constant that fixes units, and the entropic time
$t$ as a parameter designed to keep track of the accumulation of changes.
Their relation to familiar physical quantities was a matter of conjecture. But
now that we can see what role they play in the Schr\"{o}dinger equation we can
identify $m$ with the particle mass, $\hbar$ with Planck's constant, and the
entropic time $t$ with physical\ time.\footnote{Where by `physical' we mean
that it is the time $t$ that appears in the laws of physics.}

Other attempts to derive quantum theory start from an underlying, perhaps
stochastic, classical mechanics. The ED approach is different in that it does
not assume an underlying classical substrate; ED provides a derivation of
\emph{both Schr\"{o}dinger's equation and also Newton's }$F=ma$\emph{.}
Classical mechanics is recovered in the usual limits of $\hbar\rightarrow0$ or
$m\rightarrow\infty$. Indeed, writing $S=\hbar\Phi$ in eq.(\ref{SEb}) and
letting $m\rightarrow\infty$ with $S/m$ fixed leads to the classical
Hamilton-Jacobi equation%
\begin{equation}
\dot{S}+\frac{1}{2m}(\vec{\nabla}S)^{2}+V=0~, \label{HJ}%
\end{equation}
while eqs.(\ref{fluc}), (\ref{osmo}), and (\ref{curr}) give $m\vec{v}%
=\,\vec{\nabla}S$ and $\vec{u}=0$ with vanishing f{}luctuations $\left\langle
\Delta w^{a}\Delta w^{b}\right\rangle =\frac{\hbar}{m}\Delta t\,\delta
^{ab}\rightarrow0$.

\section{Measurement in ED}

In practice the measurement of position can be technically challenging because
it requires the amplification of microscopic details to a macroscopically
observable scale. However, no intrinsically quantum effects are involved: the
position of a particle has a definite, albeit unknown, value $x$ and its
probability distribution is, by construction, given by the Born rule,
$\rho(x)=|\Psi(x)|^{2}$. We can therefore assume that suitable position
detectors are in principle available. First we consider observables other than
position: how they are defined and how they are measured. Then we conclude
with a few remarks on amplification and Bayes theorem.

\subsection{Observables other than position}

The fact that the Schr\"{o}dinger equation (\ref{SE}) is linear and unitary
makes the language of Hilbert spaces particularly convenient so from now we
adopt Dirac's bra-ket notation and write $\Psi(x)=\langle x|\Psi\rangle$. For
convenience we consider the case of a particle that lives on a discrete
lattice. The generalization to a continuous space is straightforward. The
probabilities of the previously continuous positions
\begin{equation}
\rho(x)\,d^{3}x=|\langle x|\Psi\rangle|^{2}\,d^{3}x\quad\text{become}\quad
p_{i}=|\langle x_{i}|\Psi\rangle|^{2}\ ,
\end{equation}
and if the state is
\begin{equation}
|\Psi\rangle=%
{\textstyle\sum\limits_{i}}
c_{i}|x_{i}\rangle\quad\text{then}\quad p_{i}=|\langle x_{i}|\Psi\rangle
|^{2}=|c_{i}|^{2}~.
\end{equation}

Since position is the only objectively real quantity there is no reason to
define other observables except that they turn out to be convenient when
considering more complex experiments. Consider a setup in which right before
reaching the position detector the particle is subjected to additional
interactions, say magnetic fields or diffraction gratings. Suppose the
interactions in such a complex setup $\mathcal{A}$ are described by the
Schr\"{o}dinger eq.(\ref{SE}), that is, by a particular unitary evolution
$\hat{U}_{A}$. The particle will be detected with certainty at position
$|x_{i}\rangle$ provided it was initially in a state $|s_{i}\rangle$ such
that
\begin{equation}
\hat{U}_{A}|s_{i}\rangle=|x_{i}\rangle\ .
\end{equation}
Since the set $\{|x_{i}\rangle\}$ is orthonormal and complete, the
corresponding set $\{|s_{i}\rangle\}$ is also orthonormal and complete,
\begin{equation}
\langle s_{i}|s_{j}\rangle=\delta_{ij}\quad\text{and}\quad%
{\textstyle\sum\nolimits_{i}}
|s_{i}\rangle\langle s_{i}|{}=\hat{I}\ .
\end{equation}
Now consider the effect of this setup $\mathcal{A}$ on some generic initial
state vector $|\Psi\rangle$ which can always be expanded as
\begin{equation}
|\Psi\rangle=%
{\textstyle\sum\nolimits_{i}}
c_{i}|s_{i}\rangle\ ,
\end{equation}
where $c_{i}=\langle s_{i}|\Psi\rangle$ are complex coefficients. The state
$|\Psi\rangle$ will evolve according to $\hat{U}_{A}$ so that as it approaches
the position detectors the new state is
\begin{equation}
\hat{U}_{A}|\Psi\rangle=%
{\textstyle\sum\nolimits_{i}}
c_{i}\hat{U}_{A}|s_{i}\rangle=%
{\textstyle\sum\nolimits_{i}}
c_{i}|x_{i}\rangle\ .
\end{equation}
which, invoking the Born rule for position measurements, implies that the
probability of finding the particle at the position $x_{i}$ is
\begin{equation}
p_{i}=|c_{i}|^{2}=|\langle s_{i}|\Psi\rangle|^{2}\ . \label{Born a}%
\end{equation}
Thus, the probability that the particle in the initial state $|\Psi\rangle$
after going through the setup $\mathcal{A}$ is found at position $x_{i}$ is
$|c_{i}|^{2}$.

The same experiment can be described from a point of view in which the setup
$\mathcal{A}$ is a black box, a complex detector the inner workings of which
are not emphasized. The particle is detected at $|x_{i}\rangle$\emph{\ as if}
it had earlier been in the state $|s_{i}\rangle$. We can adopt a new language
and say, perhaps inappropriately, that the particle has effectively been
\textquotedblleft detected\textquotedblright\ in the state $|s_{i}\rangle$,
and therefore, the probability that the particle in state $|\Psi\rangle$ is
\textquotedblleft detected\textquotedblright\ in state $|s_{i}\rangle$ is
$|\langle s_{i}|\Psi\rangle|^{2}$ --- which reproduces Born's rule for a
generic measurement device. The shift in language is not particularly
fundamental --- it is merely a matter of convenience but we can pursue it
further and assert that the setup $\mathcal{A}$ is a complex detector that
\textquotedblleft measures\textquotedblright\ all operators of the form
\begin{equation}
\hat{A}=%
{\textstyle\sum\nolimits_{i}}
\lambda_{i}|s_{i}\rangle\langle s_{i}|
\end{equation}
where the eigenvalues $\lambda_{i}$ are arbitrary scalars.

Some remarks are in order. Note that when we say we have detected the particle
at $x_{i}$ \emph{as if} it had earlier been in state $|s_{i}\rangle$ with
eigenvalue $\lambda_{i}$ we are not implying that the particle was in the
particular state $|s_{i}\rangle$ --- this is just a figure of speech. It is in
this sense that the corresponding value $\lambda_{i}$ of the observable
$\hat{A}$ has been \textquotedblleft created by the act of
measurement\textquotedblright. To be more explicit:\ if a sentence such as
\textquotedblleft a particle has momentum $\vec{p}\,$\textquotedblright\ is
used only as a linguistic shortcut that conveys information about the wave
function before the particle enters the complex detector then, strictly
speaking, there is no such thing as the momentum of the particle. The momentum
is not an attribute of the particle; it is an attribute of the epistemic state
$\Psi(x)$.

Incidentally, note that it is not necessary that the eigenvalues of the
operator $\hat{A}$ be real --- they could be complex numbers. What is
necessary is that its eigenvectors $|s_{i}\rangle$ be orthogonal. This means
that $\hat{A}$ need not be Hermitian but its Hermitian and anti-Hermitian
parts of $\hat{A}$ must be simultaneously diagonalizable --- they must commute.

In the standard interpretation of quantum mechanics Born's rule (\ref{Born a})
is a postulate; within ED it is the natural consequence of unitary time
evolution and the fact that all measurements are ultimately position
measurements. This raises the question of whether our scheme is sufficiently
general to encompass all measurements of interest. While there is no general
answer that will address all cases --- who can, after all, even list all the
measurements that future physicists might perform? --- we can, nevertheless,
ask whether our scheme includes a sufficiently large class of interesting
measurements. How, for example, does one measure an observable for which there
is no unitary transformation mapping its eigenstates to position eigenstates?
Every case demands its own specific analysis. For example, how does one
measure the energy of a free particle? A measurement device characterized by
eigenvectors $\{|s\rangle\}$ measures all operators of the form $\hat{A}=%
{\textstyle\int}
ds\,\lambda(s)|s\rangle\,\langle s|$. Therefore the same device that measures
the momentum $\hat{p}$ of a particle (\emph{e.g.}, using a magnetic field or a
diffraction grating followed by a position detector such as a photographic
plate or a photoelectric cell) can also be used to infer the energy $\hat
{H}=\hat{p}^{2}/2m$ of a free particle.

Here is another example: It is not so easy to place a probe inside the atom,
so how does one measure the energy of an electron that is bound to an atom? In
practice the energy of the bound particle is not measured directly; instead it
is inferred from the energy of photons emitted in transitions between the
bound states. Since photons are free particles measuring their energy is not
in principle problematic. This is a special case of the general scheme in
which the system of interest and the pointer variable of an apparatus become
correlated in such a way that observation of the pointer allows one to infer a
quantity of the system. The paradigmatic example is a Stern-Gerlach experiment
in which the particle's position is the pointer variable that allows one to
infer its spin.

The difficulty with the standard von Neumann interpretation is that it is not
clear at what stage the pointer variable \textquotedblleft
collapses\textquotedblright\ and attains a definite value. This is precisely
the difficulty of principle that is resolved in the entropic approach: the
pointer variable is a position variable too and therefore always has a
definite value.

\section{Amplification}

The technical problem of amplifying microscopic details so they can become
macroscopically observable is usually handled with a detection device set up
in an initial state of unstable equilibrium. The particle of interest
activates the amplifying system by inducing a cascade reaction that leaves the
amplifier in a definite macroscopic final state described by some pointer
variable $\alpha$.

An eigenstate $|s_{i}\rangle$ evolves to a position $x_{i}$ and the goal of
the amplification process is to infer the value $x_{i}$ from the observed
value $\alpha_{r}$ of the pointer variable. The design of the device is deemed
successful when $x_{i}$ and $\alpha_{r}$ are suitably correlated and this
information is conveyed through a likelihood function $P(\alpha_{r}|x_{i})$
--- an ideal amplification device would be described by $P(\alpha_{r}%
|x_{i})=\delta_{ri}$. Inferences about $x_{i}$ follow from a standard
application of Bayes' rule,
\begin{equation}
P(x_{i}|\alpha_{r})=P(x_{i})\frac{P(\alpha_{r}|x_{i})}{P(\alpha_{r})}\ .
\end{equation}

The point of these considerations is to emphasize that there is nothing
intrinsically quantum mechanical about the amplification process. The issue is
one of appropriate selection of the information (in this case the data
$\alpha_{r}$) that happens to be relevant to a certain inference (in this case
$x_{i}$). This is, of course, a matter of design: a skilled experimentalist
will design the device so that no spurious correlations---whether quantum or
otherwise---nor any other kind of interfering noise will stand in the way of
inferring $x_{i}$.

\paragraph*{Acknowledgments}

I would like to thank D. Bartolomeo, C. Cafaro, N. Caticha, S. DiFranzo, A.
Giffin, P. Goyal, D. T. Johnson, K. Knuth, S. Nawaz, M. Reginatto, C.
Rodr\'{\i}guez, and J. Skilling for many discussions on entropy, inference and
quantum mechanics.

\end{document}